\documentstyle[multicol,aps,prl,amssymb,epsf]{revtex}
\begin{document}
\draft \title{On molecular and continuum boundary conditions for a
  miscible binary fluid}
\author{Colin Denniston and Mark O. Robbins}
\address{Department of Physics and Astronomy, The Johns Hopkins
  University, Baltimore, Maryland 21218, USA}
\date{\today} \maketitle

\begin{abstract}
We show that molecular dynamics simulations can furnish useful
boundary conditions at a solid surface bounding a two-component fluid.
In contrast to some previous reports, convective-diffusive flow is
consistent with continuum equations down to atomic scales.  However, concentration gradients can produce flow without
viscous dissipation that is inconsistent with the commonly used Navier slip
condition.  Also, differential wetting of the two components coupled
to a concentration gradient can drive convective flows that could be
used to make nano-pumps or motors.
\end{abstract}
\begin{multicols}{2}

Hydrodynamic theories of flow past a solid surface need to assume boundary
conditions for the fluid velocities at the surface.
Such interfacial behavior is often very difficult to access experimentally.   
Recent simulation studies of fluids have revealed a range of
boundary conditions for single component fluids and related
them to the microscopic interactions \cite{TR90,BB99}.
However, there are many problems where the appropriate boundary conditions
are still in doubt.
These include flow near a moving contact line,
the liquid crystal order parameter
in the presence of flow,
and convective-diffusive flow of miscible fluids.

In a recent letter \cite{KB98} Koplik and Banavar studied two fluids
that individually obeyed a ``no-slip'' boundary condition at the
surface.  In a mixture of these fluids, they found 
that the velocities of each component still vanished at the wall.
They concluded that the boundary condition accompanying the
convective-diffusive transport of a binary fluid mixture demands that
the velocities of the two species be equal at the wall.
It was subsequently pointed out \cite{GB99} that this boundary condition
contradicts Fick's law of molecular diffusion, at least in its usual
form where the diffusion coefficient is assumed to be position
independent.  Two possibilities were suggested.  The first
\cite{GB99}, that the convective flows studied in Ref.~\cite{KB98}
were large enough to mask the effect of diffusion.  The
second, that the continuum equations need to be modified near the
surface in order to obtain consistent results \cite{GB99,KB99}. 

In an attempt to resolve this issue, Brenner and Ganesan \cite{BG00}
undertook a singular perturbation analysis. A continuum inner region with a
refined form of Fick's law on which the no-slip condition applies
for both components was asymptotically matched to an outer region which
follows the standard form of Fick's law.  As the details of the
refined form of Fick's law are still somewhat ambiguous, Brenner and
Ganesan conclude that ``simulations purporting to derive''
boundary conditions ``by direct probing of those
molecules near to the wall will necessarily give rise to erroneous
macroscale conclusions''.  Essentially, they argue that the
simulations are useless since they do not give the information
necessary to match up to the known equations for bulk fluids.  If
true, this would be a very disappointing situation.   

In this paper we examine flow boundary conditions for binary fluid 
mixtures using molecular dynamics simulations in the regime where
diffusive flow is either dominant or on the same scale as the
convective flow.  We find bulk flow consistent with Fick's law
within {\it one} molecular diameter from the wall, and appropriate
boundary conditions for the known equations of bulk fluids are readily
obtained.  If concentration gradients are present, the individual
velocities of the two species are not equal at 
the wall.  More surprisingly, we find that density gradients can lead
to a net flow without viscous dissipation!  
We also find that the coupling of wetting potentials with
concentration gradients can give rise to Marangoni-like convective
flow that could be used to make nano-pumps or nano-motors.  

We consider a mixture of two types of molecules, labeled $a$ and $b$,
in a slit geometry similar to that used for a single fluid in
Ref.~\cite{TR90}.  There are two walls, at $z=0$ and $z=L$, and
periodic boundary conditions in the $x-y$ plane.  The walls contain
type $w$ atoms fixed to lattice sites forming two $(111)$ layers of an fcc
surface.  The interactions between atoms of type $i$ and $j$ separated
by a distance $r$ are modeled using a Lennard-Jones (LJ) potential,
$
V_{ij}(r)=4 \epsilon_{ij} \left[ (\sigma_{ij}/r)^{12}-(\sigma_{ij}/r)^6\right],
$
where $\epsilon_{ij}$
specifies the interaction energy and $\sigma_{ij}$ the interaction
length.  Their averages are denoted by $\epsilon$ and $\sigma$,
respectively.  A
characteristic time scale is given by $\tau=\sigma(m/\epsilon)^{1/2}$,
where $m$ is the average of the molecular masses $m_i$.  Unless
specified, the force is truncated at $r_c=2.2 \sigma$ to improve
computational efficiency.

To obtain a steady-state concentration gradient we artificially
construct a direction dependent osmotic membrane at $x=0$.  The
membrane is less than one $\sigma$ thick, and preferentially
transmits atoms of type $a$ from left to right and conversely for atoms
of type $b$.  In a typical simulation we wait $5000 \tau$ for the
steady-state concentration gradient to establish itself, and then
average over a subsequent $15000$ to $150000 \tau$ to collect
data. Note that these time scales are more than an order of magnitude
greater than in Ref.~\cite{KB98}, thus allowing much greater sensitivity.
Most channels studied had $L=16.4\sigma$, but we have also
found similar results for systems twice as wide.

The diffusive flux ${\bf J}_j$ of particles of type $j$,
\begin{equation} 
{\bf J}_j \equiv \rho_j({\bf v}_j-{\bf v}),
\label{defnflux}
\end{equation} 
is defined \cite{dGM84} relative to the barycentric, or mass averaged,
velocity ${\bf v}\equiv (\sum_i \rho_i {\bf v}_i)/\rho$, where $\rho_i$ is the
position-dependent mass density of species $i$, and
$\rho \equiv\sum_j \rho_j$ is the total mass density. Fick's
law relates the diffusive flux to the gradient of the concentration
$c_j\equiv\rho_j/\rho$\cite{dGM84}, 
\begin{equation}
{\bf J}_j   = -\rho D \nabla c_j,
\label{Fick}
\end{equation}
where $D$ is the diffusion constant.  The results of
Refs. \cite{KB98,GB99,KB99,BG00} suggest that there may be a
macroscopic layer where the diffusion constant depends on proximity to
the wall. 

\begin{figure}
\centerline{\epsfxsize=3.2in
\epsffile{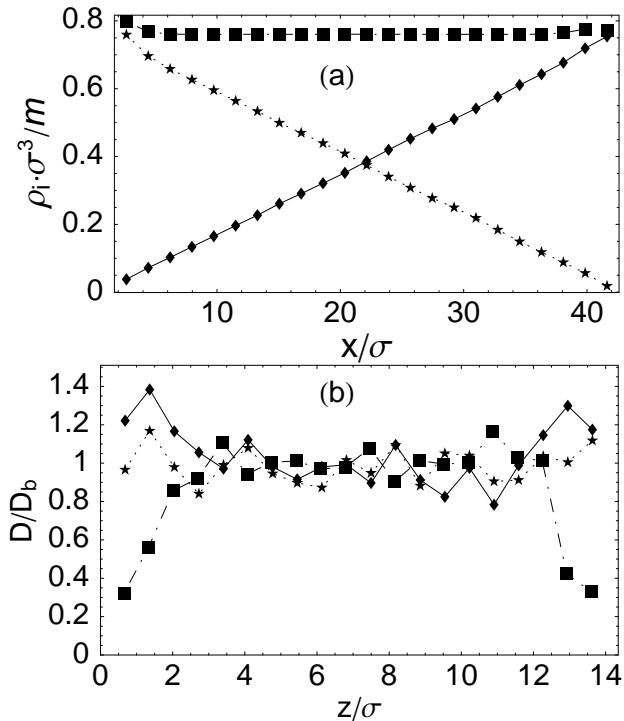}}
\vskip 0.02true cm
\caption{(a) Density of fluid particles of type $a$ ($\blacklozenge$) and $b$
  ($\star$), and total density ($\blacksquare$).  The osmotic membrane
  is located at $x=0$.  The periodic cell dimensions along the
  $x$ and $y$-directions are $42.5\sigma$ and $12.3\sigma$, respectively.
  (b) Diffusion constant for $\epsilon_{wf}/\epsilon=1.0$
  ($\blacksquare$), $0.3$ ($\star$), and $0.1$ ($\blacklozenge$).  We
  normalize by the bulk diffusion constant, $D_b=0.068 \sigma^2/\tau$,
  measured in a system without 
  walls.  Away from the osmotic membrane, the value of $D$ is
  independent of $x$ and $y$.}
\label{AequalB}
\end{figure}

For simplicity, we first consider
a system where $\sigma_{ij}= \sigma$ for all
interactions and $m_a=m_b= m$.  All fluid-fluid
$\epsilon_{ij}=\epsilon$, and the wall-fluid coupling
$\epsilon_{wf}\equiv\epsilon_{wa}=\epsilon_{wb}$ is varied.  
In this case entropy is the only
driving force.  Figure~\ref{AequalB}(a) shows the resulting steady
state densities $\rho_j$ of the two species.  Both vary linearly and
the total density is constant throughout the system.  Since $a$ and
$b$ particles have equal and 
opposite concentration gradients we find that their fluxes are equal
and opposite and ${\bf v}=0$.
If the difference between individual velocities vanished at the wall,
as in Ref. 4, both fluxes would have to vanish there.
Figure \ref{AequalB}(b) illustrates that the fluxes 
may actually increase at the wall.
To eliminate variations with $x$ we plot $D=v_{x,j} (c_j/\partial_x c_j)$ as
a function of height $z$ for three different wall-fluid interaction
strengths.
The diffusion constant, and thus the flux, goes down near the wall for
$\epsilon_{wf}/\epsilon=1.0$, remains constant for
$\epsilon_{wf}/\epsilon=0.3$, and goes up for
$\epsilon_{wf}/\epsilon=0.1$.
Thus, for $\epsilon_{wf}/\epsilon=0.1$, the difference between the
individual velocities of the two species is actually highest at the wall.
The value of $D$ is normalized by the bulk
diffusion constant which was calculated for a system where the walls were
replaced by periodic boundary conditions in the $z$ direction.  As can
be seen, there is no statistical difference from bulk
diffusion at distances more than about one molecular diameter from the
wall.  Thus, in contrast to References~\cite{KB98,GB99,KB99,BG00}, there is
no need to supplement Fick's law with a {\it continuum} boundary layer which
obeys a modified constitutive relation.

Earlier work\cite{TR90,BB99} on single component LJ fluids 
is well described by the Navier boundary condition,
which assumes that the velocity difference between solid and fluid
is proportional to the viscous shear stress.
For our geometry this implies
\begin{equation}
v_x|_{wall} = L_s \partial_z v_x |_{wall},
\label{eq:navier}
\end{equation}
where $L_s$ is called the slip length.
A larger value of $L_s$ implies less drag at the interface,
and we find that it correlates with greater diffusion at the wall.
In most cases, the slip length has atomic dimensions
\cite{TR90,BB99},
and it is less than $2\sigma$
for single fluids with the walls considered here.

\begin{figure}
\centerline{\epsfxsize=3.4in
\epsffile{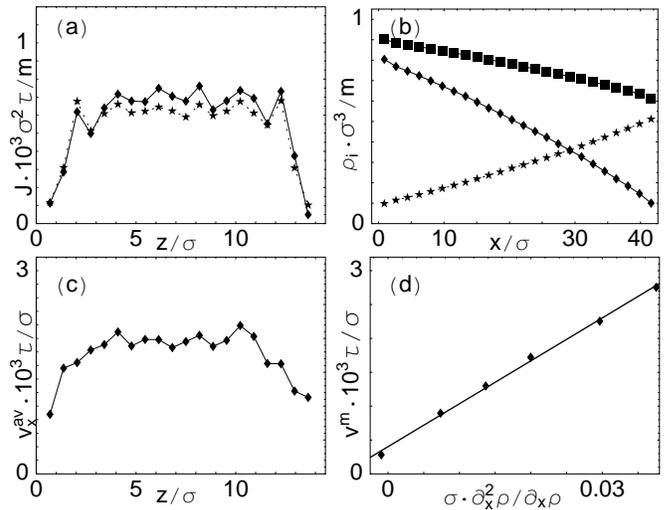}}
\vskip 0.02true cm
\caption{(a) The observed mass flux $J=\rho v_x$ ($\blacklozenge$) and
  that predicted $J=-D\partial_x \rho$ ($\star$) by Eq.~(\ref{vdiff}) for
  ${\bf v}^m=0$. 
  (b) Mass density of fluid particles of type $a$ ($\blacklozenge$) and $b$
  ($\star$) and total density ($\blacksquare$).
  (c) The mass flux normalized by the mean density
  $\rho_0$, $v_x^{av}=\rho v_x/\rho_0$.
  (d) Mean molar velocity versus $\partial_x^2 \rho/\partial_x
  \rho$ for different osmotic walls.  The line is a linear fit giving
  $D=0.067\pm0.004 \sigma^2/\tau$.}
\label{Am125Bm075}
\end{figure}

The Navier condition presupposes 
that in the purely diffusive case, where $\partial_z v_x=0$,
the barycentric velocity vanishes identically.
However, Fig.~\ref{Am125Bm075}(a) shows that this is not 
necessarily the case.
For this simulation the masses were changed to
$m_a=0.75 m$ and $m_b=1.25 m$ and $\epsilon_{wf}/\epsilon=1.0$.
We find that the {\it number} densities $N_i$ are the same, within
statistical errors, as those of Fig.~\ref{AequalB}(a)
and the mean velocities of each species are also unchanged.
However, due to the mass difference, there is now a mass density
gradient and a small net mass flux ${\bf J} = \rho {\bf v}$
along the channel (Fig. \ref{Am125Bm075}(a)).
One might wonder if the Navier condition still applies, but
with a very small velocity gradient and very large slip length.
However, both single and two-component simulations of purely
convective flow for the same parameters
yield ``stick'' boundary conditions, i.e. the effective
``wall'' position is inside the fluid and $L_s$ is essentially zero.
Moreover, there is no measurable viscous stress
and the entire flux profile follows from a general
continuum relation that we now describe.

For the example of Fig.~\ref{Am125Bm075}(a) the
mean molar velocity ${\bf v}^m\equiv(\sum_i N_i {\bf v}_i)/N$ vanishes,
where $N_i$ is the molar density of component $i$ and $N\equiv\sum_i N_i$.
Using the definition of the diffusive flux (Eq.~(\ref{defnflux})),
and Fick's law (Eq.~(\ref{Fick})) one can show that 
\begin{equation}
{\bf v}={\bf v}^m+D\nabla \left[\ln (N/\rho) \right].
\label{vdiff}
\end{equation}
Thus the barycentric and molar velocities differ when there is a
diffusive flow driven by a gradient in a total density ($N$ or $\rho$).  
In the case considered in Fig.~\ref{AequalB},
the total densities are constant and both ${\bf v}$
and ${\bf v}^m$ vanish everywhere for purely diffusive flow.
For the case of Fig.~\ref{Am125Bm075}(a), ${\bf v}^m$ vanishes
and $N$ is essentially constant.
Thus Eq. \ref{vdiff} implies ${\bf J}=\rho {\bf v} = - D \nabla \rho$.
The dashed line in Fig.~\ref{Am125Bm075}(a) shows that this
expression reproduces the observed mass flux.
Note that the height dependence comes entirely from the variation
in $D$ that is also evident in Fig. \ref{AequalB}(b).
If smaller values of $\epsilon_{wf}/\epsilon$ are used, the
flux may actually increase at the wall.
This leads to different signs on the two sides of Eq. \ref{eq:navier}!

The observed failure of the Navier condition has important consequences.
Numerous arguments related to the theory of diffusive flow, such as those in
Ref.~\cite{CB95}, rely on the Navier condition to enforce a vanishing
barycentric velocity in the absence of diffusion.
The conclusions of these works need to be reexamined.
One may wonder whether the correct boundary condition for purely
diffusive flow is the identical vanishing of some other mean velocity
such as ${\bf v}^m$. However, it is easy to construct exceptions to such
ansatzes. The difficulty with such boundary conditions is most readily
seen by considering steady state situations.
Steady state implies that $\partial_t \rho=\partial_t N_i=0$ and the
corresponding continuity equations then imply that
\begin{equation}
\partial_\alpha (\rho v_\alpha)=\partial_\alpha (N v_\alpha^m)=0.
\label{steadystate}
\end{equation}
Multiplying Eq. (\ref{vdiff}) by $N$ and taking the
divergence yields
\begin{equation}
\partial_\alpha (N v_\alpha)=\partial_\alpha (N v_\alpha^m)+
    \partial_\alpha\left(N D\partial_\alpha \left[\ln (N/\rho) \right]\right).
\end{equation}
Rewriting $\partial_\alpha (N v_\alpha)$ as $\partial_\alpha\left[ (N/\rho)
(\rho v_\alpha)\right]$ and making use of Eq.~(\ref{steadystate}), one
can easily obtain 
\begin{equation}
 v_\alpha \partial_\alpha(N/\rho)=\frac{1}{\rho}\partial_\alpha\left(N
   D \partial_\alpha \left[\ln (N/\rho) \right]\right).
\label{vrelate}
\end{equation}
An analogous relation for the mean molar velocity can easily be derived,
\begin{equation}
 v_\alpha^m \partial_\alpha(\rho/N)=\frac{1}{N}\partial_\alpha\left(\rho D
  \partial_\alpha \left[\ln (N/\rho) \right]\right).
\label{vmrelate}
\end{equation}
Thus the continuum relations themselves place constraints on mean
velocities. Any additional requirement that a mean velocity vanish
imposes constraints on the gradients of the density.  While these may
be consistent with some situations, they will in general over-specify
the problem.

We have tested the above equations with simulations using a range
of atom sizes and masses.
To illustrate their application,
we use the parameters for Fig. \ref{Am125Bm075}(a),
but adjust the probability of reflection at the
osmotic wall to produce different net average mass flows down the
channel.  Typical examples of the resulting steady state mass density
and flow profiles for $\epsilon_{wf}/\epsilon=0.3$
are shown in Fig.~\ref{Am125Bm075}(b) and (c).
Neither the barycentric nor the mean molar velocity vanishes.  Indeed, they
are independent of height at distances more than a molecular
diameter from the wall.  
We find that the mean molar density $N$ is independent of $x$, and
that only the $x$ components of the velocity are nonzero.
Thus Eq. (\ref{vmrelate}) simplifies to 
$v_x^m=D \partial_x^2 \rho/\partial_x \rho$.
The spatially averaged $v_x^m$ versus $\partial_x^2
\rho/\partial_x \rho$ for several different osmotic boundary
conditions is shown in Fig.~\ref{Am125Bm075}(d).  We find a good fit
to a straight line whose slope $D=0.067\pm0.004$ agrees with the value
$D=0.070\pm0.004$ found directly from Fick's law
(Eq.~(\ref{Fick})).

It is clear from the above examples that for multicomponent fluids the
Navier condition must be modified by subtracting the diffusive flow from the
left hand side of Eq.(\ref{eq:navier}).  We find that simulations of a variety
of convective-diffusive flows can then be fit with a common value of
the slip length\cite{DRx}.

\begin{figure}
\centerline{\epsfxsize=3.4in
\epsffile{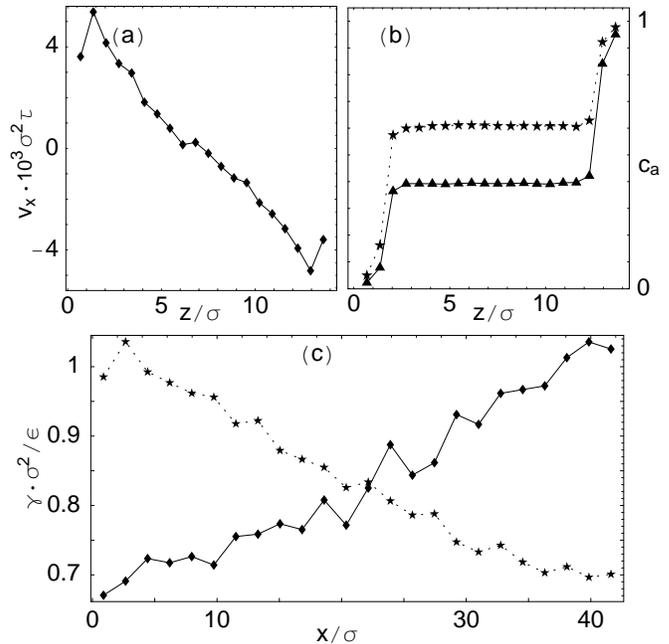}}
\vskip 0.02true cm
\caption{(a) Barycentric velocity down the channel for a case where
  the wall at $z=0$ preferentially wets $a$ and the other wall
  preferentially wets $b$.
  (b) Mass fraction $c_a=\rho_a/\rho$ versus height evaluated at planes $1/4$
  ($\blacktriangle$) and $3/4$ ($\star$) of the way along the channel
  in  the $x$-direction.  
  (c) Surface tension along the top ($\blacklozenge$) and bottom
  ($\star$) walls.}
\label{wdiff}
\end{figure}

So far we have examined only neutral wetting where both $a$ and $b$ have the
same interaction with the walls.  In many realistic cases, one
component will preferentially wet the wall.  To examine this situation
we return to the case where the two particles are indistinguishable
except for their labels.  For particles $a$ ($b$) we now change the wall-fluid
interaction at the top (bottom) wall so that it is purely repulsive by
truncating the potential at its minimum rather than at $2.2\sigma$.
Recall that previously for this fluid we found that all average
velocities were zero. Figure~\ref{wdiff}(a) shows the average velocity
produced by changing the wetting properties.  One sees a remarkably
strong shear flow ($\sim$m/s) with significant ``slip'' at the
stationary walls. This slip has the wrong sign for the Navier
condition and its magnitude is inconsistent with results for
convective flows. 

The driving force for the flow in Fig.~\ref{wdiff}(a) comes from the externally
imposed concentration gradient and the variation of the
interfacial free energy of the walls with concentration.
Figure~\ref{wdiff}(b) shows the mass fraction $c_a$ at different
cross-sections along the channel.  As can be seen from the figure,
along the walls the system prefers the more strongly wetting fluid species.
The difference between the concentration
at the wall and in the center of the channel increases with $x$.   This
increase leads to a rise in interfacial tension.  Values of the
surface tension $\gamma$, calculated using the mechanical
definition of Kirkwood and Buff \cite{KB49}, are shown in
Fig.~\ref{wdiff}(c).   
The boundary condition relating the shear stress on the wall
$\sigma_{xz}|_w$ to the viscous stress in the fluid $\sigma_{xz}|_f$ 
 is analogous to that used for Marangoni flow at a two fluid
boundary \cite{marangoni}:
\begin{equation}
\sigma_{xz}|_w-\sigma_{xz}|_f= \partial_x \gamma.
\end{equation}
It can be shown \cite{DRx} that there is also a velocity discontinuity
given by an integral of the Marangoni stress over the interfacial
region that produces the slip in Fig.~\ref{wdiff}(a). This can be
incorporated into a generalized Navier slip condition by adding a
source term proportional to $\partial_x \gamma$. 
These boundary conditions determine the net stress on the
wall. The result is consistent with stresses measured directly in our
simulations, and ranges from 50 to 75\% of $\partial_x \gamma$
($\sim$MPa) for the cases we have studied.
 This force could be used
to drive a nano-motor.  We have also constructed systems where both
the top and bottom walls prefer the same component.  In this case, one
generates Poiseuille-like flow which could be used in a nano-pump.

In conclusion, we find that the flow boundary condition for
convective-diffusive flow is not one of equal velocities for all species.
In addition, average velocities do not, in general, vanish at the wall
in the absence of viscous stress.  Diffusive mass
transport can contribute to a significant average velocity at the
wall.  Further, the presence of concentration gradients along the wall
in the general case of non-neutral wetting can result in significant
Marangoni-type forces which drive convective flow.  These effects 
should be readily applicable to the design of new micro-fluidic
devices and may be relevant to the function of numerous biological
systems.

This material is based upon work supported by the National Science
Foundation under Grant No. 0083286.  We thank Intel Corporation for
donating the workstations used for our simulations, which were
performed using LAMMPS from Sandia National Laboratories.

\vfill
\end{multicols}
\end{document}